\documentclass[12pt]{article}
\usepackage{epsfig,a4}
\usepackage{cite}
 \advance\oddsidemargin by -1in
 \advance\evensidemargin
 by -1in
 \marginparsep
 0.4cm
 \advance\topmargin by -0cm
 \makeindex
 \def\gsim{\mathrel{\rlap{\lower4pt\hbox{\hskip1pt$\sim$}}
 \raise1pt\hbox{$>$}}}

 \newcommand\la{\langle}
 \newcommand\ra{\rangle}
 \newcommand\beq{\begin{equation}}
 \newcommand\noi{\noindent}
 \newcommand\eeq{\end{equation}}
 \newcommand\beqn{\begin{eqnarray}}
 \newcommand\eeqn{\end{eqnarray}}

\def\fm{\,\mbox{fm}}
\def\GeV{\,\mbox{GeV}}

\def\lsim{\mathrel{\rlap{\lower4pt\hbox{\hskip1pt$\sim$}}
    \raise1pt\hbox{$<$}}}         
\def\gsim{\mathrel{\rlap{\lower4pt\hbox{\hskip1pt$\sim$}}
    \raise1pt\hbox{$>$}}}         
\def\lg{l^g_c}
\def\lpi{L^\pi_c}

\def\fm{\,\mbox{fm}}
\def\GeV{\,\mbox{GeV}}

\def\s0{\sigma_0(s)}

\begin{document}
\date{}

\title{\bf Jet lag effect and leading hadron production}

\maketitle

\begin{center}

\vspace*{-1.5cm}
 {\large B.Z.~Kopeliovich$^{1-3}$, H.-J.~Pirner$^2$, I.K.~Potashnikova$^1$
and Ivan~Schmidt$^1$}
 \\[0.5cm]
{$^1$Departamento de F\'{\i}sica y Centro de Estudios
Subat\'omicos,
Universidad T\'ecnica Federico Santa Mar\'{\i}a,
Casilla 110-V,
Valpara\'\i so, Chile}\\[0.2cm]
$^2${\sl Institut f\"ur Theoretische Physik der Universit\"at,
Philosophenweg 19, 69120
Heidelberg, Germany}\\[0.2cm]
$^3${\sl Joint Institute for Nuclear Research, Dubna, Russia}
\end{center}

\vspace{1cm}

\begin{abstract}

We propose a solution for the long standing puzzle of a too steeply
falling fragmentation function for a quark fragmenting into a pion,
calculated by Berger \cite{berger} in the Born approximation.
Contrary to the simple anticipation that gluon resummation worsens
the problem, we find good agreement with data. Higher quark Fock
states slow down the quark, an effect which we call jet lag. It can
be also expressed in terms of vacuum energy loss. As a result, the
space-time development of the jet shrinks and the $z$-dependence
becomes flatter than in the Born approximation. The space-time
pattern is also of great importance for in-medium hadronization.

 \end{abstract}

\vspace{1cm}



\section{Leading hadrons in Born approximation}

We are interested here in the production of leading pions which
carry a major fraction of the momentum of a highly virtual quark
originating from a hard reaction. The Born graph for the
perturbative fragmentation $q\to \pi q$ is shown in Fig.~1a,
 \begin{figure}[htp]
\centerline{
  \scalebox{0.55}{\includegraphics{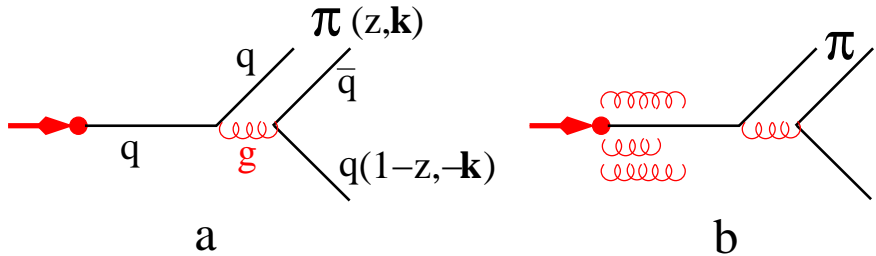}}
 }
\caption{\label{berger}\sl
 {\bf a:} Berger mechanism \cite{berger} of leading pion production
in Born approximation. {\bf b:} a high Fock component of the quark
emerging from a hard reaction and producing a pion with a higher momentum
fraction $\tilde z>z$ than measured experimentally.
 }
 \end{figure}
and the corresponding fragmentation function was calculated in
\cite{berger},
 \beq
\frac{\partial D^{(Born)}_{\pi/q}(z)}{\partial k^2}\propto
\frac{(1-z)^2}{k^4}\,,
\label{10}
 \eeq where $k$ and $z$ are the transverse and fractional longitudinal
momenta of the pion. This expression is derived under the conditions
$1-z\ll1$ and $k^2\ll Q^2$, where $Q^2$ is the scale of the hard
reaction. We neglect higher twist terms \cite{berger,pg}, which are
specific for deep-inelastic scattering (DIS).

The fragmentation function (FF) Eq.~(\ref{10}) is in an apparent
contradiction to data, since it falls towards $z=1$ much steeper
than is known from phenomenological fits (e.g. see \cite{kkp}), and
even the inclusion of higher order correction does not seem to fix
the problem. Moreover, at first glance gluon radiation should worsen
the situation, producing even more suppression at $z\to1$ because of
energy sharing.

Nevertheless, we demonstrate below that the effect of jet lag (JL),
i.e. the effect that comes from the fact that higher Fock states
retard the quark, substantially changes the space-time pattern of
jet development. The JL cuts off contributions with long coherence
time in pion production and makes the $z$-dependence less steep.

One can rewrite (\ref{10}) in terms of the coherence length of pion
radiation,
 \beq
\lpi=\frac{2Ez(1-z)}{k^2+z^2m_q^2+(1-z)m_\pi^2}\,,
\label{20}
 \eeq
 where $E$ is the jet energy, and $m_q$ is the quark mass which may be
treated as an effective infrared cutoff. Then, the Born approximation
takes the form,
\beq
\frac{\partial
D^{(Born)}_{\pi/q}(z)}{\partial \lpi}
\propto (1-z)\,.
\label{30}
 \eeq
 Thus, the production of the leading pion is homogeneously distributed over
distance, from the point of jet origin up to the maximal distance
$(\lpi)_{max}=2E(1-z)/zm_q^2$.

Integrating (\ref{30}) over $\lpi$ up to $(\lpi)_{max}$ we recover the
$(1-z)^2$ dependence of Eq.~(\ref{10}). Now we understand where the extra
power of $(1-z)$ comes from: it is generated by the shrinkage of the
coherence pathlength for $z\to1$. This is the source of the too steep fall
off of the Born term Eq.~(\ref{10}) in the FF.

\section{Jet lag effect and the fragmentation function}

The color field of a quark originated from a hard reaction
(high-$p_T$, DIS, $e^+e^-$, etc.) is stripped off, and gluon
radiation from the initial state generates the scale dependence of
the quark structure function of the incoming hadron (if any).
Therefore the quark originated from such a hard process is bare,
lacking a color field up to transverse frequencies $q\lsim Q$. Then
the quark starts regenerating its field by radiating gluons, i.e.
forming a jet. This can be described by means of an expansion of the
initial "bare" quark over Fock states containing a physical quark
and different number of physical gluons with different momenta, as
is illustrated in Fig.~\ref{berger}b.  Originally this is a coherent
wave packet equivalent to a single bare quark $|q\ra$. However,
different components have different invariant masses and start
gaining relative phase shifts as function of time. As a result, the
wave packet is losing coherence and gluons are radiated in
accordance with their coherence times.

Notice that the Born expression (\ref{10}) corresponds to the lowest Fock
components relevant to this process, just a bare quark, $|q\ra$, and a
quark accompanied by a pion, $|q\pi\ra$. In this case the initial quark
momentum and the pion fractional momentum $z$ in (\ref{10}) are the
observables (at least in $e^+e^-$ or SIDIS).

An important observation is that the quark in higher Fock states
carries only a fraction of the full momentum of the wave packet. At
the same time, the pion momentum is an observable and is fixed.
Therefore, one should redefine the fractional momentum of the pion
convoluting the fragmentation function Eq.~(\ref{10}) with the quark
momentum distribution within different Fock states,
 \beq
\frac{\partial D_{q/\pi}(z)}
{\partial \lpi}=
\left\la\frac{\partial D_{q/\pi}(z)}
{\partial \lpi}\right\ra_x=
\frac{
\sum\limits_i C^q_i
\int\limits_z^1 d x\,
\frac{\partial D^{(Born)}_{q/\pi}(z/x)}
{\partial \lpi}
F^i_q(x)\,\Theta(\lpi-l_c^i)}
{\sum\limits_i C^q_i
\int\limits_z^1 d x\,
F^i_q(x)\,\Theta(\lpi-l_c^i)}
\,.
\label{50}
 \eeq
 Here $F^i_q(x)$ is the fractional momentum distribution function of a
physical quark in the $i$-th Fock component of the initial bare quark.
Such a component contributes to (\ref{50}) only if it lost coherence with
the rest of the wave packet. This is taken into account in (\ref{50}) by
means of the step function, where $l^i_c$ is the coherence length for this
Fock state. We sum in (\ref{50}) over different Fock states with proper
weight factors $C^q_i$.

Thus, the inclusion of higher Fock states results in a retarding of
the quark, an effect which we call jet lag (JL). This effect plays a
key role in shaping the quark fragmentation function for leading
hadrons. Due to JL the variable of the Born FF in (\ref{50})
increases, $z\Rightarrow z/x$, causing a suppression.
 Then the convolution Eq.~(\ref{50}) leads to the following
modification of the Born fragmentation function
Eq.~(\ref{30}),
 \beq
\frac{\partial D_{q/\pi}(z)}
{\partial \lpi} \propto
1-\tilde z\,,
\label{57}
 \eeq
 where
 \beq
\tilde z =
\left\la {z\over x}\right\ra =
z\left(1+\frac{\Delta E}{E}\right) +
O\left[z(1-z)^2\right]\,.
\label{55}
 \eeq
 Here we made use of the limiting behavior at $1-z\ll1$ we are interested
in. The fractional energy loss of the quark is related to the energy
carried by other partons within those Fock components which have lost
coherence one the pathlength $\lpi$,
 \beq
\frac{\Delta E(\lpi)}{E}=
\la1-x(\lpi)\ra=
\frac{\sum\limits_i C^q_i
\int_z^1 d x\,(1-x)\,F^i_q(x)\,
\Theta(\lpi-l_c^i)}
{\sum\limits_i C^q_i
\int_z^1 d x\,F^i_q(x)\,
\Theta(\lpi-l_c^i)}\,.
\label{60}
 \eeq

Notice that in the above expressions we implicitly assume also
integration on the other kinematic variables related to the
participating partons.

\subsection{Gluon bremsstrahlung}

A part of energy loss related to radiation of gluons can be evaluated
perturbatively. For this purpose we replace $F^i_q(x)$ in (\ref{50}) and
(\ref{60}) by the gluon number distribution \cite{gb},
 \beq
\frac{dn_g}{d\alpha dk^2} = \frac{2\alpha_s(k^2)}{3\pi}\,
\frac{1+(1-\alpha)^2}{\alpha\,k^2}\,,
\label{80}
 \eeq
 where $\alpha=1-x$ is the fraction of the total energy carried by the
radiated gluon, and the fractional momentum of the recoil quark is
$x$. In the numerator we added the splitting function of the DGLAP
equations, although it is a small corrections, since $\alpha\ll1$.

Then, the perturbative vacuum energy loss for gluon radiation reads
\cite{feri,knp,knph},
 \beq
\Delta E_{pert}(L) =
E\int\limits_{\lambda^2}^{Q^2} dk^2
\int\limits_{k/2E}^{1}
d\alpha\,\alpha\,\frac{dn_g}{dk^2d\alpha}\
\Theta(L-\lg)\Theta\left(1-z-\alpha-\frac{k^2}
{4\alpha E^2}\right)
\label{90}
 \eeq
 Here the soft cutoff $\lambda$ is fixed at $\lambda=0.7\GeV$. The latter
choice is dictated by data (see in \cite{kst2,spots}) demonstrating a
rather large primordial transverse momentum of gluons.

The first step-function in (\ref{90}) restricts the radiation time
of gluons,
 \beq
\lg=\frac{2E\alpha(1-\alpha)}{k^2}\,,
\label{92}
 \eeq
 contributing to the quark energy loss along the pathlength $L$.
The second step-function in (\ref{90}) takes care of energy
conservation, namely, none of the gluons can have energy,
$\omega=\alpha E+k^2/4\alpha E$, larger than $E(1-z)$.

One can rewrite Eq.~(\ref{80}) at $\alpha\ll1$ as a distribution of
gluon number over the radiation length and fractional momentum,
 \beq
\frac{dn_g}{d\lg d\alpha} = \frac{4\alpha_s(\mu^2)}{3\pi}\,
\frac{1}{\lg\,\alpha}\,,
\label{95}
 \eeq
 where the scale in the running QCD coupling is
$\mu^2=2E\alpha(1-\alpha)/\lg$.
Then, Eq.~(\ref{90}) takes the form,
 \beq
\Delta E_{pert}(L) =
E\int\limits_{1/Q}^{l_{max}} dl
\int\limits_{(2El)^{-1}}^{1}
d\alpha\,\alpha\,\frac{dn_g}{dld\alpha}\,
\Theta\left(1-z-\alpha-\frac{1-\alpha}{2lE}\right)
\,,
\label{100}
 \eeq
 where the upper limit of integration over $l$ is given by the maximal
value, $l_{max}={\rm min}\{L,\ E/2\lambda^2\}$.

\begin{figure}[htp]
\centerline{
  \scalebox{0.45}{\includegraphics{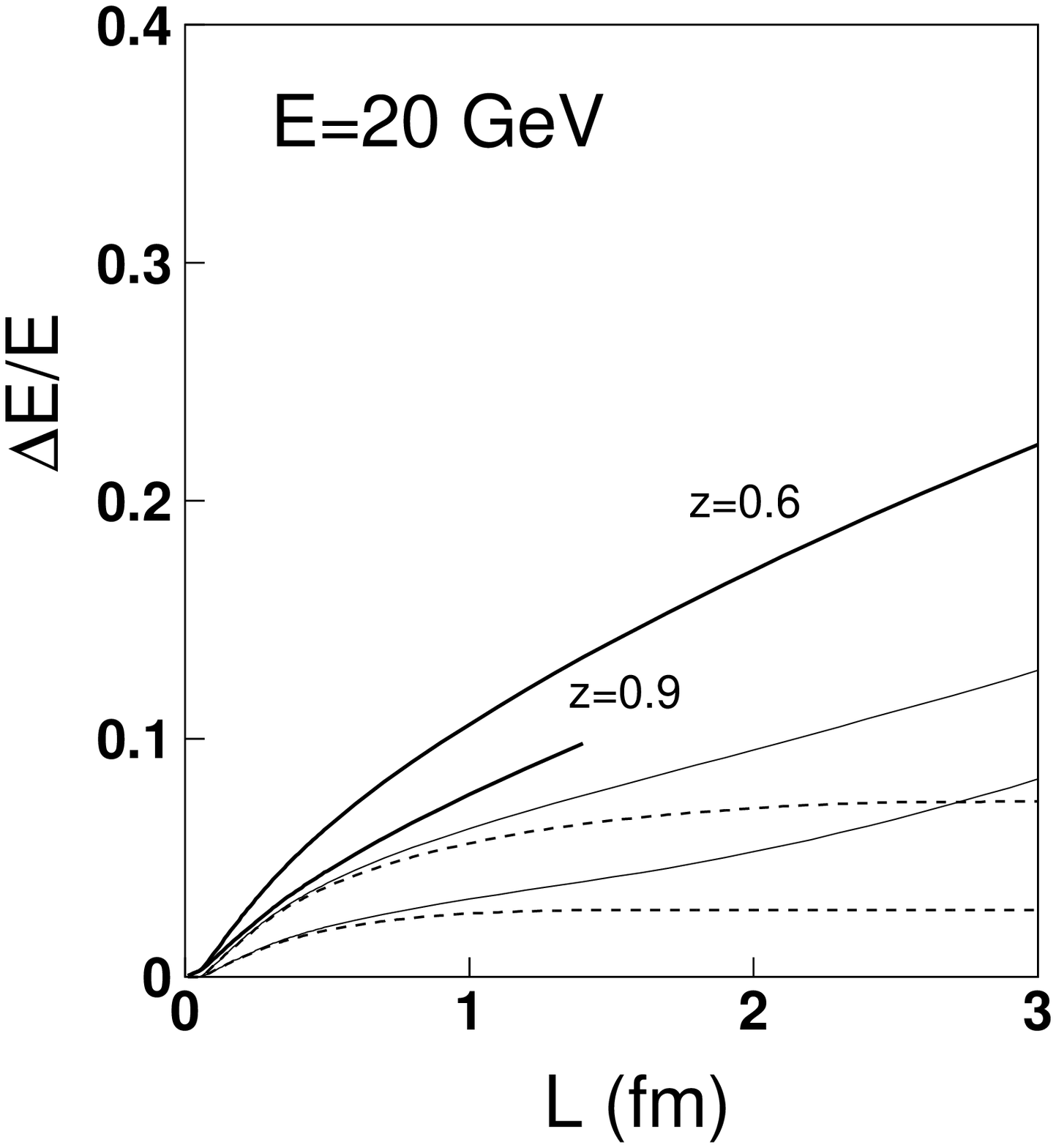}}
  \scalebox{0.45}{\includegraphics{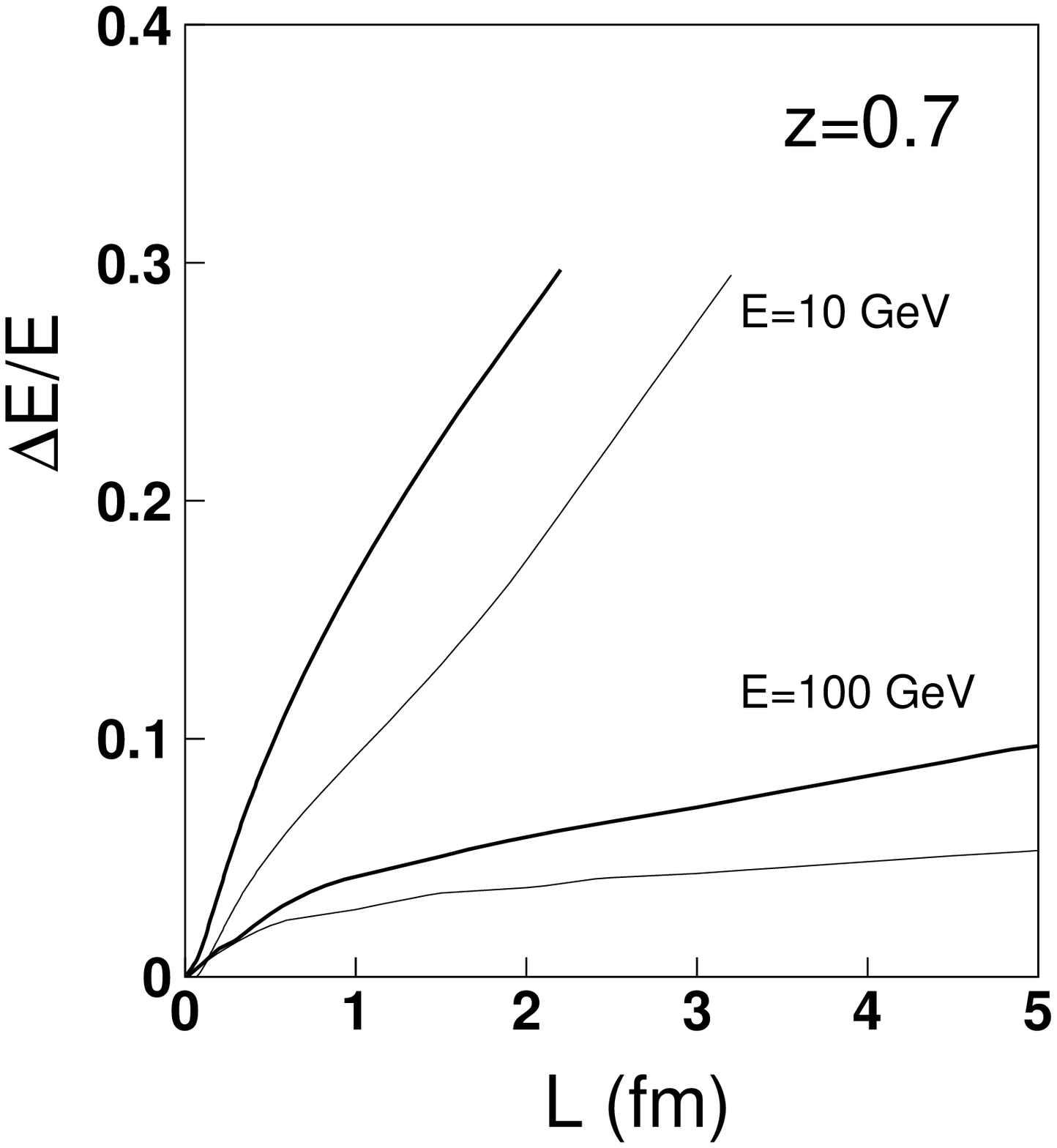}}
 }
\caption{\label{e-loss}\sl
 Fractional energy loss by a quark with $Q=E$ producing a hadron
with fractional momentum $z$ is depicted by solid curves as function of
distance $l$. Dashed curves show the perturbative contribution.  The
nonperturbative part for thick and thin solid curves is calculated with two
models, STRING-I and STRING-II respectively.
 }
 \end{figure}

An example of $L$-dependence of fractional energy loss calculated with 
Eq.~(\ref{100}) for $E=Q=20\GeV$ is shown in Fig.~\ref{e-loss} by dashed 
curve (left panel).
 
We see from Eq.~(\ref{90}) that the rate of the perturbative energy loss
ceases at $L>E/2\lambda^2$, since no gluons are radiated any more. Of
course propagation of a free quark is unphysical and the effects of
confinement at a scale softer than $\lambda$ must be introduced.

\subsection{Sudakov suppression}

As far as we imposed a ban for radiation of gluons with energy
$\omega>(1-z)E$ in (\ref{90}),(\ref{100}), this restriction leads to a
Sudakov type suppression factor,
 \beq
S(L,z)=\exp\left[-\la n_g(L,z)\ra\right]\,,
\label{120}
 \eeq
 where $\la n_g(L,z)\ra$ is the mean number of nonradiated gluons,
 \beq
 \la n_g(L,z)\ra=\int\limits_{1/Q}^{l_{max}} dl
\int\limits_{(2El)^{-1}}^{1}
d\alpha\,\frac{dn_g}{dld\alpha}\,
\Theta\left(\alpha+\frac{1-\alpha}{2lE}-1+z\right)
\,.
\label{130}
 \eeq

\begin{figure}[htb]
\centerline{
  \scalebox{0.45}{\includegraphics{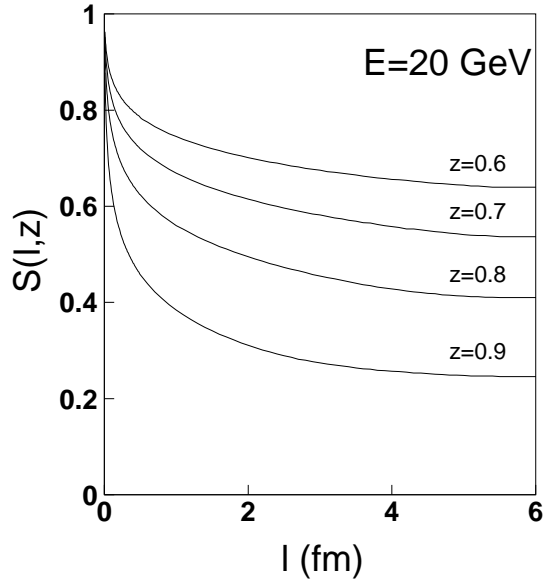}}
 }
\caption{\label{sudakov}\sl
Sudakov suppression caused by a ban for radiation of gluons with
fractional energy higher than $1-z$. Calculations are done for a jet with
$E=Q=20\GeV$
 }
 \end{figure}

The results are illustrated in Fig.~\ref{sudakov}, at $E=Q=20\GeV$
and for different values of $z$.

\subsection{Higher twist nonperturbative effects also contribute}

At long distances $L>E/2\lambda^2$, after completing restoring its
field, the quark does not radiate any more, and then the energy loss
may have only a nonperturbative origin. We assume that a soft quark
develops a string (color flux tube \cite{cnn}), which leads to a
constant rate of energy loss \cite{kn,k},
 \beq
\left.\frac{dE(L>E/2\lambda^2)}{dL}
\right|_{string} = -\kappa\,,
\label{105}
 \eeq
 where the string tension is taken at its static value $\kappa=1\GeV/\fm$,
given by the slope of Regge trajectories and by calculations on the
lattice.

At shorter distances, $L<E/2\lambda^2$, the nonperturbative energy
loss proceeds along with the perturbative one. The way how it is
introduced is the most uncertain and model dependent part of the
calculation, since we have no good knowledge of the relevant
dynamics.  Nevertheless, this is a higher twist effect, and the
related uncertainties tend to vanish at high $Q^2$.

To see the range of this uncertainty we consider two models for
nonperturbative energy loss.

{\it STRING-I}: we assume the constant rate of energy loss Eq.~(\ref{105})
to be valid at all distance from the origin.

{\it STRING-II}: keeping the same rate of energy loss
Eq.~(\ref{105}) at long distances $L>E/2\lambda^2$, we reduce and
make time-dependent the rate of nonperturbative energy loss at
$L<E/2\lambda^2$. Indeed, the original bare quark whose field has
been stripped off, cannot produce any color flux at the origin, and
starts developing a flux tube only during restoration of its field.
We assume that the transverse area of the color flux follows the
transverse size of the restored field of the quark, which receives
contributions from all the gluons radiated during quark propagation
through the pathlength $L$ ($\lg<L<E/2\lambda^2$),
 \beq
\la r^2\ra \sim \left\la{1\over k^2}\right\ra
\propto \frac{L}{2E}\,.
\label{107}
 \eeq
 Therefore, in this scenario the transverse area of the color flux formed by a quark rises
linearly with the pathlength of the quark.

In the MIT bag model the energy of a tube comes from two
contributions, the bag term and the energy of the electric color
field \cite{cnn}. The first one is proportional to the transverse
area times the bag constant, while the second contribution has
inverse dependence on the tube area. Equilibrium corresponds to
equal contributions of these two terms. If such a tube fluctuates to
a smaller transverse dimension $r$, the second term rises as
$1/r^2$, and so does the string tension. Nevertheless, this is
probably true for a stationary tube when the total flux of electric
color field is independent of the transverse size of the tube and is
equal to the color charge of the quark. However, a quark with a
stripped field produces a color flux only at the transverse
distances where the field is already restored. Therefore, both terms
in the energy of a flux tube produced by a bare quark are reduced by
the same factor $\la r^2\ra/a^2$, where $\la r^2\ra$ is given by
(\ref{107}), and $a$ is the transverse size of a stationary tube.

The parameter $a$ in the stochastic vacuum model
\cite{dosch,nachtmann}, has the meaning of a gluon correlation
radius, calculated on the lattice \cite{pisa} as $a=0.3-0.35\fm$.
This value turns out to be in a good accord with our infrared cutoff
in Eq.~(\ref{90}), $a\approx 1/\lambda$.

Thus, the mean transverse dimension squared of the flux rises
linearly with $L$ from a tiny value $r^2\sim Q^{-2}$ up to the
stationary value $r^2=a^2\approx1/\lambda^2$. Correspondingly, the
effective string tension rises $\propto r^2(L)$ with a coefficient
dependent on the QCD gluon condensate \cite{nachtmann},
 \beq
\kappa_{eff}(L)=\frac{32\pi k}{81}\,
\left\la {\alpha_s\over\pi}G_{\mu\nu}^a(x)
G^{\mu\nu a}(0)\right\ra\,r^2(L)\,,
\label{108}
 \eeq
 where $k=0.74$.

Summarizing, the nonperturbative energy loss rises linearly with
pathlength during gluon radiation and restoration of the color field of
the quark.
The energy loss rate approaches its maximum value Eq.~(\ref{105})
at the maximal length available for radiation, $L_{max}=E/2\lambda^2$,
so we can write,
 \beq
\left.\frac{dE(L<E/2\lambda^2)}{dL}
\right|_{string} = -\frac{2\lambda^2}{E}\,
L\,\kappa\,.
\label{109}
 \eeq

Although we believe that this model String-II is more realistic than the
previous one, in what follows we perform calculations with both models to
see the range of theoretical uncertainty.

We add the two sources of energy loss, the perturbative gluon radiation
and the string contribution.  Fig.~\ref{e-loss} shows an example of length
dependence of the fractional energy loss, by a quark with $E=Q=20\GeV$ and
different fractional momenta $z$ of produced pions (left panel), and at
different energies, but fixed $z=0.7$ (right panel). Notice that curves
stop when all energy available for gluon radiation is exhausted, i.e.
$\tilde z\to1$. The smaller is $(1-z)$, the earlier this happens.

\subsection{Jet lag modified fragmentation function}

Now we are in a position to calculate the quark-to-pion
fragmentation function, based on Berger's result \cite{berger}
obtained in Born approximation, and corrected for gluon resummation.
Gluon radiation results in the JL effect, Eqs.~(\ref{50}),
(\ref{60}), since the pion momentum fraction should be redefined
relative to the retarded quark Eq.~(\ref{55}). Then we arrive at the
$L$-dependent fragmentation function,
 \beq
\frac{\partial D_{\pi/q}(z)}{\partial \lpi}\propto
(1-\tilde z)\,S(\lpi,z)\ .
\label{140}
 \eeq

\begin{figure}[htb]
\centerline{
  \scalebox{0.45}{\includegraphics{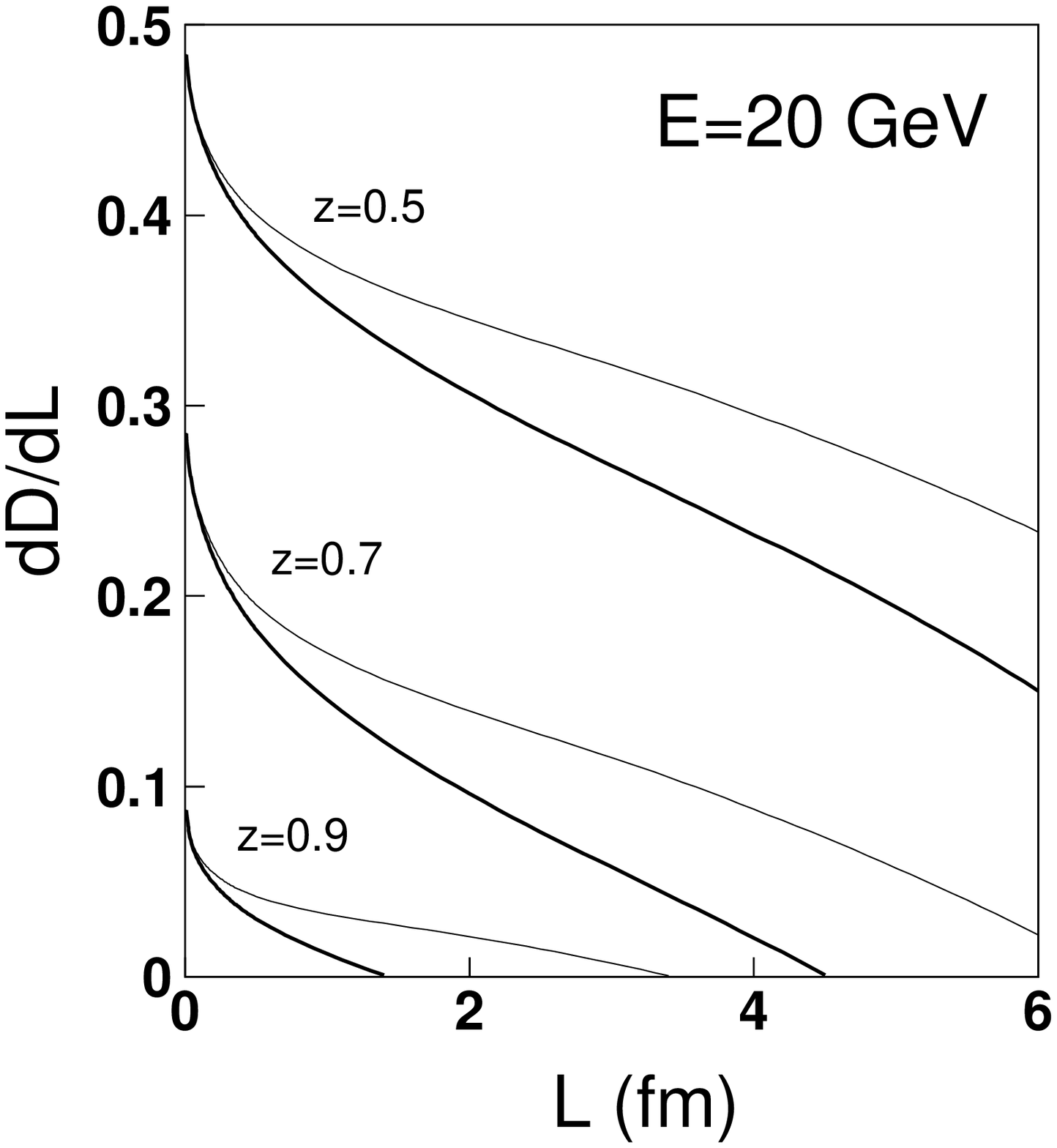}}
  \scalebox{0.45}{\includegraphics{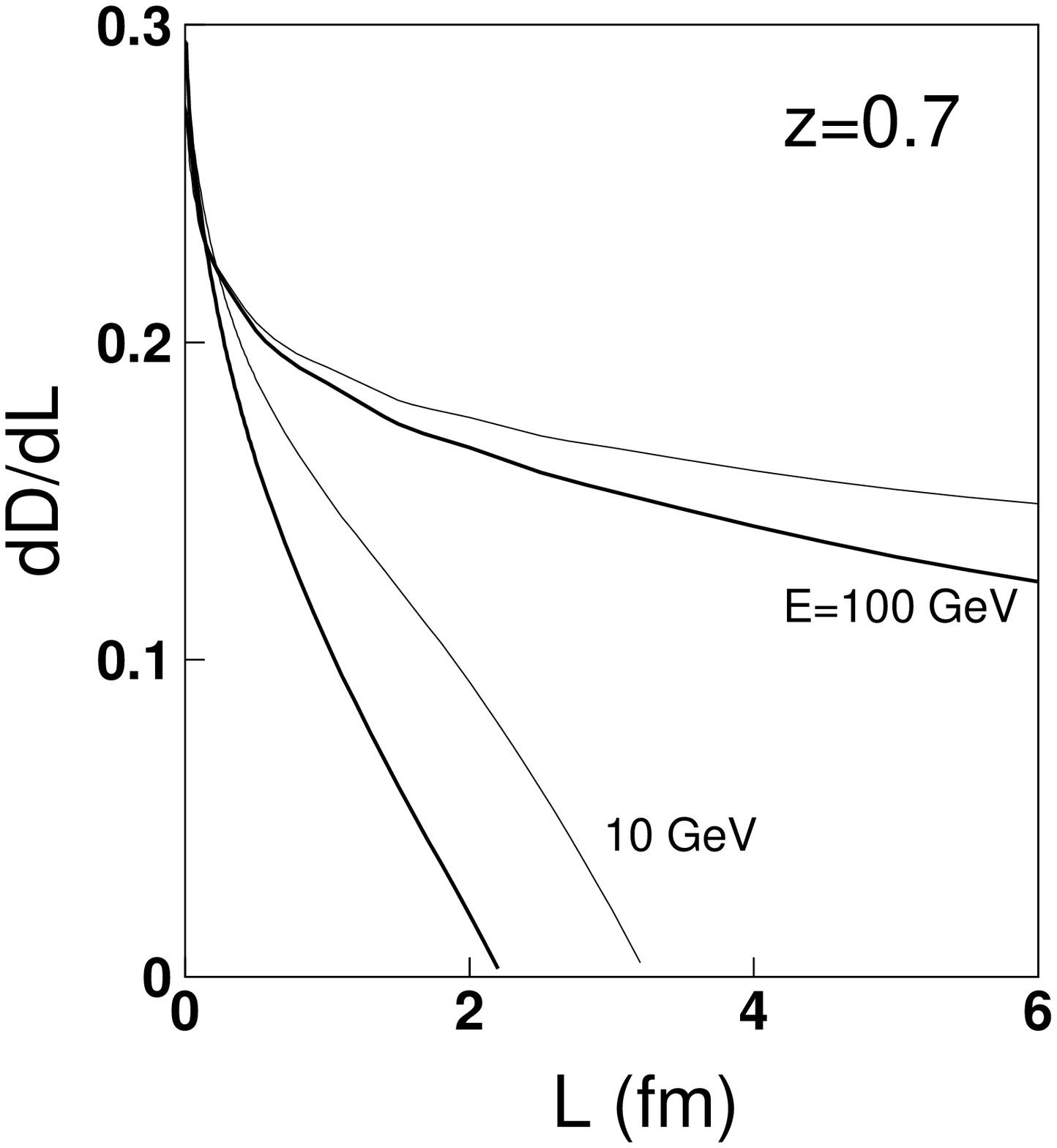}}
 }
 \caption{\label{dd-dl}\sl
 The pion production rate as function of length calculated according to
Eq.~(\ref{140}). Solid and thin curves correspond to the nonperturbative
part calculated with models STRING-I and STRING-II respectively. {\rm
left:} $E=Q=20\GeV$, $z=0.5,\ 0.7,\ 0.9$.  {\rm right:} $z=0.7$ and
$E=Q=10,\ 100\GeV$. The overall normalization is arbitrary.
 }
 \end{figure}

The JL effect and Sudakov factor suppress long distances in pion
production. The $L$-distribution at $E=Q=20\GeV$ is depicted in the
left panel of Fig.~\ref{dd-dl}, for $z=0.5,\ 0.7,\ 0.9$.
 The right panel of Fig.~\ref{dd-dl} shows the energy dependence of
the $L$-distribution at $z=0.7$. Apparently the production length of
leading pions is rather short, even at high energies.

\begin{figure}[htb]
\centerline{
  \scalebox{0.45}{\includegraphics{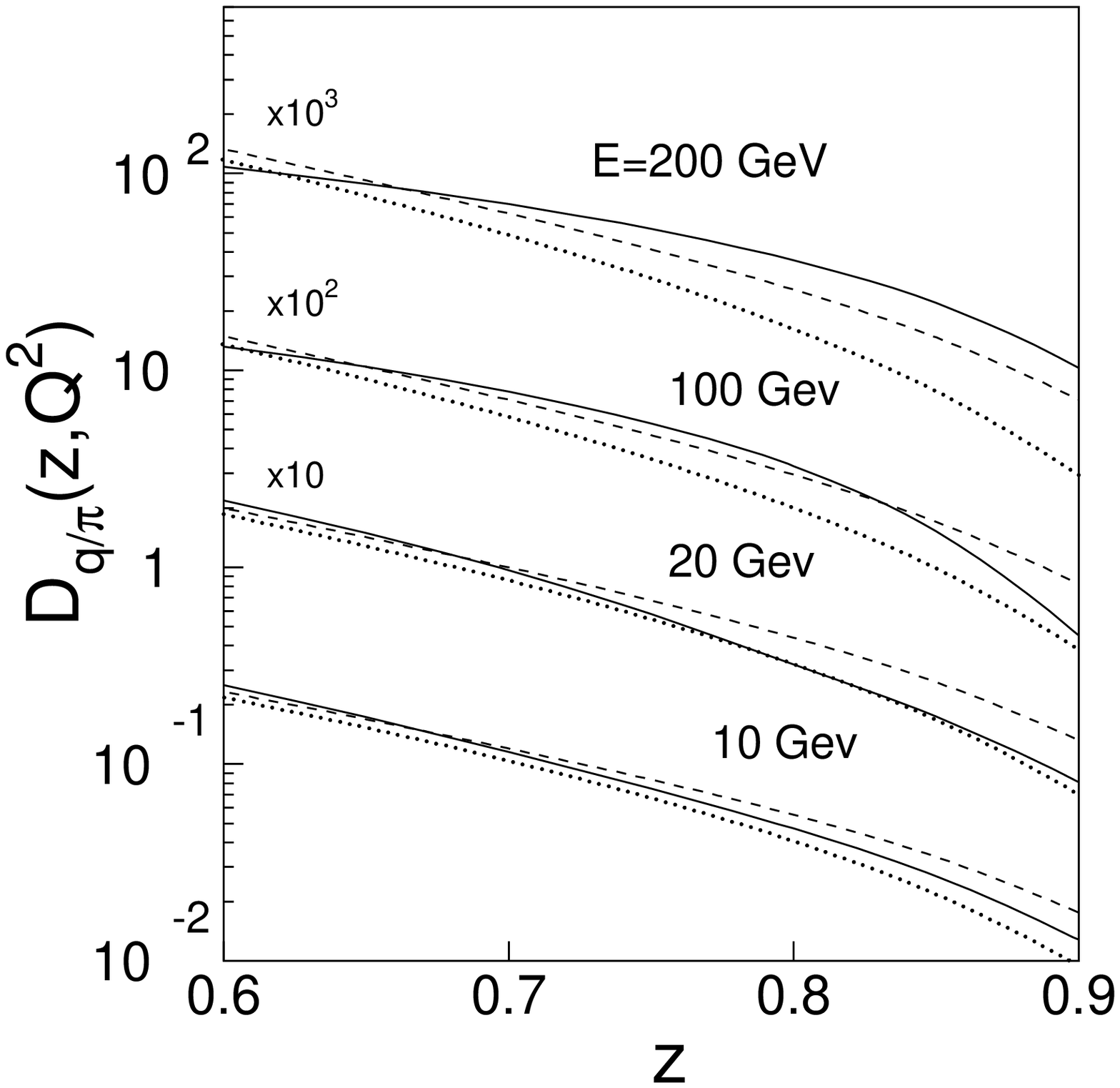}}
  \scalebox{0.45}{\includegraphics{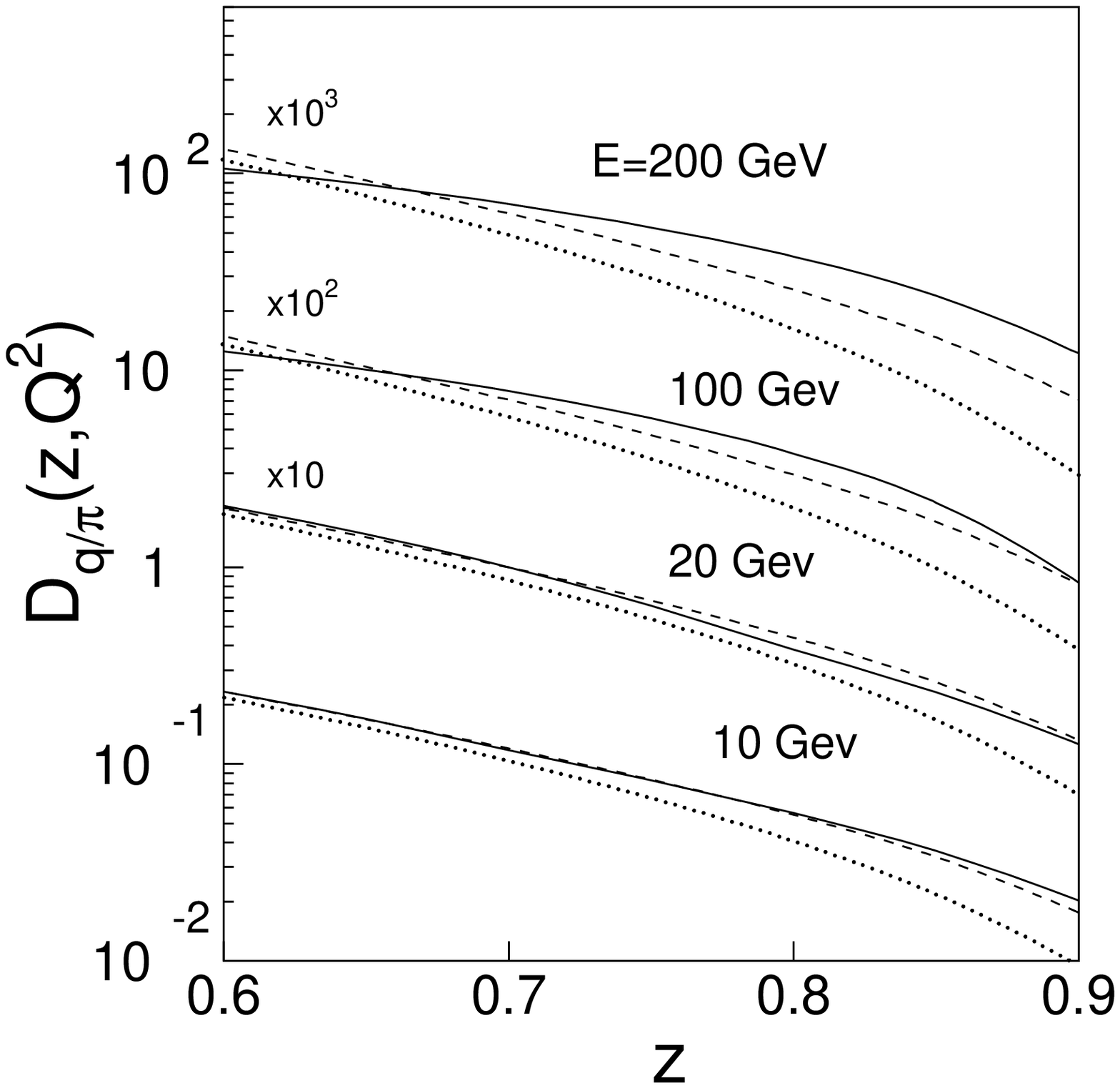}}
 }
\caption{\label{ff-kkp}\sl
 Comparison of our modeled FF (solid curves) with the phenomenological
ones \cite{kkp} (dashed) and \cite{bkk} (dotted), at scales $E=Q=5,\ 20,\
100,\ 200\GeV$. Each curve is rescaled by factor $10$ compared to the
lower one.  The nonperturbative part is calculated with either the model
STRING-I (left), or STRING-II (right).}
 \end{figure}

Integrating the distribution function Eq.~(\ref{140}) over $\lpi$ we
arrive at the fragmentation function $D_{q/\pi}(z,Q^2)$, which is
compared in Fig.~\ref{ff-kkp} with two popular parametrizations
fitted to data, KKP \cite{kkp} and BKK \cite{bkk}.
 Since our fragmentation function is valid only at large $z$ and is not
normalized, we fix the normalization adjusting it to the KKP results
at $z=0.6-0.8$. We calculated the nonperturbative part with either
constant (STRING-I) or rising (STRING-II) rates of energy loss.

We observe a rather good agreement between our calculated and the
phenomenological fragmentation functions, with deviations which are
similar to the differences between the two phenomenological FF
depicted in the figure. Notice that according to \cite{kkp} the
results of fits are not trustable at $z>0.8$ due to lack of data.

Thus, after inclusion of higher order corrections the quadratic
$(1-z)^2$ behavior of the Born approximation Eq.~(\ref{10}) is
replaced by a less steep dependence which complies well with data.
This could happen only if the interval of accessible coherence
lengths for pion production does not shrink $\propto(1-z)$ anymore.
This seems to be in contradiction with the usually anticipated
behavior \cite{kn,bg,k,knp,pir,knph},
 \beq
L_c(z)\approx\frac{E(1-z)}{\la|dE/dL|\ra}\,,
\label{150}
 \eeq
 which is dictated by energy conservation. The rate of radiative energy
loss is known to be constant \cite{feri} like in the string model
\cite{kn,k}, and then the coherence length Eq.~([\ref{140}) should
be $\propto(1-z)$.

However, the energy conservation restrictions should be imposed to the
rate of radiative energy loss as well \cite{knp,knph}. This was done above
in Eq.~(\ref{90}) by introducing the second step function.  The
corresponding rate of energy loss at small $1-z<LQ^2/2E$ reads,
 \beq
\frac{dE}{dL} = -\frac{4\alpha_s E}{3\pi L}\,
(1-z)\,,
\label{160}
 \eeq
 where we fix $\alpha_s$ (only here) at the scale $\mu^2=2Ez(1-z)/L$.

Thus, in the limit $z\to1$ the rate of energy loss is small $\propto(1-z)$
and the interval of coherence length Eq.~(\ref{150}) does not shrink.
A combination of radiation and nonperturbative sources of energy loss
results in in a $z$-dependence which lies in between of liner and
quadratic $1-z$ behaviors, as is demonstrated in Fig.~\ref{ff-kkp}.

\section{Summary and outlook}

The Born approximation for leading pions result in a FF,
Eq.~(\ref{10}), which drops too steeply at $z\to1$. We complemented
this result with a space-time evolution pattern, and found that
pions are produced along a long path whose length rises with jet
energy. Long distances turn out to be responsible for the extra
power of $1-z$ in the FF.

Gluon radiation leads to vacuum energy loss, which considerably
reduces the pathlength of the quark. In terms of Fock state
decomposition the effective value of the pion fractional momentum,
Eq.~(\ref{55}), is larger, since the presence of other partons
(gluons) is retarding the quark, an effect named JL.

Our central result, the final expression for the FF, given in
Eq.~(\ref{140}), also includes the Sudakov suppression factor which
causes more shrinkage of the pion production length, and which is
found to be rather short as is demonstrated in Fig.~\ref{dd-dl}.
Integrating over $\lpi$ we arrived at a FF which agrees quite well
with phenomenological ones fitted to data, as is shown in
Fig.~\ref{ff-kkp}.

There is still much work to be done:
\begin{itemize}

\item Understanding the space-time pattern of hadron production is crucial
for the calculation of medium modification of the FF. The present
results can be applied to hadron attenuation in SIDIS off nuclei,
and also to hadron quenching in heavy ion collisions \cite{cern}.

\item Transverse momentum distributions are always more difficult to
calculate than the integrated FF. Nevertheless, the current approach
has a predictive power for the $k_T$-distribution as well.
The results will be published elsewhere.

\item
Quark fragmentation to leading heavier flavor mesons can be
calculated as well. In this case a mass correction in the Born
approximation energy denominator needs to be done. We also leave
this for further study.

\end{itemize}

\bigskip

\noi {\bf Acknowledgments:} We are thankful to Dima Antonov and Sasha
Tarasov for many informative discussions. This work was supported in part
by Fondecyt (Chile) grants 1050519 and 1050589, and by DFG (Germany)
grant PI182/3-1.

\end{document}